# Universal Scaling Law for the Condensation Energy, U, Across a Broad Range of Superconductor Classes

J. S. Kim, G. N. Tam, and G. R. Stewart

Department of Physics, University of Florida, Gainesville, FL 32611

**Abstract:** One of the goals in understanding any new class of superconductors is to search for commonalities with other known superconductors. The present work investigates the superconducting condensation energy, U, in the iron based superconductors (IBS), and compares their U with a broad range of other distinct classes of superconductor, including conventional BCS elements and compounds and the unconventional heavy Fermion, $Sr_2RuO_4$, $Li_{0.1}ZrNCl$, $\kappa$-$(BEDT$-$TTF)_2Cu(NCS)_2$ and optimally doped cuprate superconductors. Surprisingly, both the magnitude and $T_c$ dependence ($U \propto T_c^{3.4\pm0.2}$) of U are – contrary to the previously observed behavior of the specific heat discontinuity at $T_c$, $\Delta C$, – quite similar in the IBS and BCS materials for $T_c > 1.4$ K. In contrast, the heavy Fermion superconductors' U vs $T_c$ are strongly (up to a factor of 100) enhanced *above* the IBS/BCS while the cuprate superconductors' U are strongly (factor of 8) *reduced*. However, scaling of U with the specific heat $\gamma$ (or $\Delta C$) brings all the superconductors investigated onto one *universal* dependence upon $T_c$. This apparent universal scaling $U/\gamma \propto T_c^2$ for all superconductor classes investigated, both weak and strong coupled and both conventional and unconventional, links together extremely disparate behaviors over almost seven orders of magnitude for U and almost three orders of magnitude for $T_c$. Since U has not yet been explicitly calculated beyond the weak coupling limit, the present results can help direct theoretical efforts into the medium and strong coupling regimes.

## I. Introduction

Not long after the discovery of iron based superconductivity by Kamihara et al. in 2008[1], Bud'ko, Ni and Canfield (BNC) noted[2] in a collection of superconducting doped $BaFe_2As_2$ samples that the discontinuity, $\Delta C$, in the specific heat at $T_c$ varied approximately as the cube of the superconducting transition temperature, $T_c^3$. This global correlation has since been confirmed[3-5] in a wide range of iron based superconductors (IBS), including not just the Ba 122 structure but also in Eu and Sr based 122 as well as in the 11 and 111 materials. In those confirmations, it was found that the exponent $\alpha$ in $\Delta C \propto T^\alpha$ could vary between 2.5 and 3, depending on annealing/sample quality. Making this correlation of some fundamental importance was the observation[4] that BCS superconductors obey $\Delta C \propto T_c^\beta$, $\beta \sim 1.8$, thus the BNC correlation reveals an *intrinsic* (still under theoretical discussion[6-8]) difference between IBS – where pairing is believed to be mediated by spin fluctuations - and BCS electron-phonon coupled superconductors.

Rather than $\Delta C$, the current work focuses on the superconducting condensation energy, U. Once a material becomes superconducting, the electrons enter into the more ordered Cooper paired state. This results in a lower entropy with decreasing temperature than in the normal state. As required of a second order phase transition, the two entropies are equal at $T_c$: $S_{normal}(T_c) = S_{superconducting}(T_c)$. The integral between T=0 and $T_c$ of $\{S_n(T) - S_s(T)\}$ is equal to the superconducting condensation energy, U, or the energy reduction achieved by condensing into the more ordered superconducting ground state see, e. g., Tinkham, Introduction to Superconductivity.[9] To help visualize this during the following discussion, Fig. 1 offers an example of this in Ta, $T_c$=4.48 K, where the lattice phonon contribution to both $S_n(T)$ and $S_s(T)$ has been previously subtracted, leaving just the electronic entropies. (Since the phonon contribution to the specific heat is the same in both the superconducting and normal state, the lattice contribution would be eliminated in the difference $S_n(T) - S_s(T)$ in any case.)

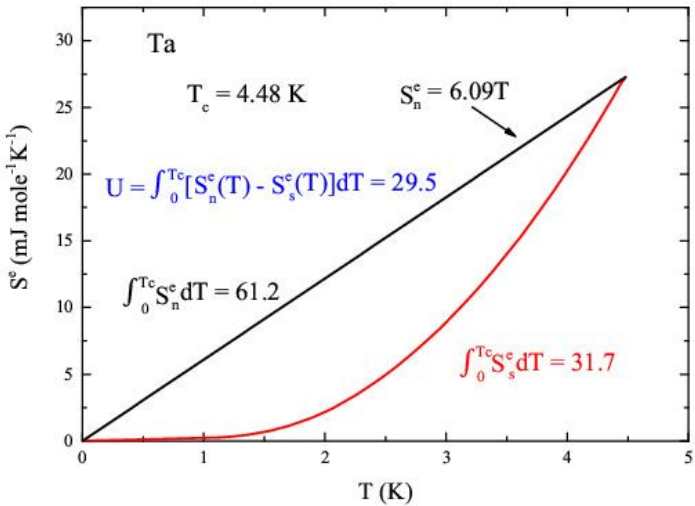

Fig. 1 (color online) Electronic entropy vs temperature, T, in both the normal state (black line), $S_n^e$ – where the normal state specific heat data, $C_n=\gamma T+\beta T^3$, were measured by the application of a field of 3000 G to suppress superconductivity – and in the superconducting state (red line), $S_s^e$ for superconducting Ta[10]. The normal state electronic entropy is just the electronic specific heat coefficient $\gamma$, 6.09 mJ/molK$^2$, times T. Units for S are mJ/molK and for U are mJ/mol. Data were measured down to ~0.8 K.

For type I superconductors, where the (easily) measured upper critical field, $H_{c1}$, is just simply also the thermodynamic critical field, $H_{c0}$, the relation $U=H_{c0}^2/8\pi$ can be used to check the condensation energy calculated as sketched in Fig. 1. For the example given in Fig. 1, Ta, $U= [H_{c0}^2/8\pi]$*molar mass/density (to get U in units of mJ/mol instead of erg/cm$^3$) and using $H_{c0}$=829 G results in U=30.2 mJ/mol – rather good agreement with the entropy integral method. In the case of Pb and Hg, where the soft lattices makes the difference between the superconducting and normal state specific heats quite small and thus the accurate determination of U difficult, calculating U via $H_{c0}^2/8\pi$ is more accurate.

The consideration of the condensation energy to gain insights into the mechanism of superconductivity – the subject of the current work - has a long history in the study of cuprate superconductivity, see refs. 11-17. There the effects of strong coupling were calculated to play a decisive role, and the details of the pairing mechanism have been shown[15] to be less important. As will be discussed more fully below, comparison of U even just amongst the cuprates themselves is difficult, due to the large changes in U (see ref. 18) with relatively small changes in doping and $T_c$ due to the formation of the pseudogap just below optimal doping. Thus the current work considers just a representative subset of the cuprates <u>at optimal doping</u>, without any pseudogap present.

Strong coupling has long been known[19] to decrease U, e. g. in Pb by 25%, vs the BCS weak coupling result $U=N(0)\Delta^2/2$, where N(0) is the electron density of states at the Fermi energy and $\Delta$ is the energy gap. In heavy Fermion superconductors, there exists one result for U in $CeCu_2Si_2$[20], but overall discussing U in heavy Fermions to understand the superconductivity has not been emphasized before now, perhaps partly due to a problem something like that of the pseudogap in cuprates. Specifically, heavy Fermion materials have an enormous range of $C_{electronic}/T$ values (which as discussed above enter into the calculations of entropy and thence on to U) which could, a priori, imply that no intercomparison of the U in heavy Fermion superconductors amongst themselves, or with other superconducting classes, is possible. As we will see, although heavy Fermion values of U do indeed exhibit a large variation when plotted vs $T_c$, with proper scaling intercomparison is indeed possible.

The present work calculates the superconducting condensation energy U using literature data (see Table 1 for the complete list of superconductors with references) for the specific heats (except for Pb and Hg and elements with $T_c$<1.4 K where such data are mostly lacking) of 1.) BCS superconductors (24 elements and three A15 compounds to extend the $T_c$ range), 2.) 12 heavy Fermion superconductors, 3.) four optimally doped cuprates, 4.) data on six different compositions, 11.6 K ≤ $T_c$ ≤ 24.6 K, of our own annealed $Ba(Fe_{1-x}Co_x)_2As_2$ single crystal samples (partially discussed in refs. 3 and 21) as well as 5.) literature data for a broad range of other IBS (FeSe, FeTe$_{0.57}$Se$_{0.43}$, LiFeAs, Ba$_{0.65}$Na$_{0.35}$-Fe$_2$As$_2$, and Ba$_{0.6}$K$_{0.4}$Fe$_2$As$_2$.) In addition, the two band gap BCS compound MgB$_2$, the organic κ-(BEDT-TTF)$_2$Cu(NCS)$_2$, the electron doped layered metal nitride halide

Li$_{0.1}$ZrNCl as well as the p-wave superconductor Sr$_2$RuO$_4$ are included, see values in Table 1. Initially, the investigation of the behavior of U was limited to the IBS and BCS superconductors for T$_c$>1.4 K, in the same vein as the ΔC work. However, initial results motivated broadening the comparison to a broader selection of superconductor classes and to lower T$_c$ values for the elements.

We therefore present a comparison of these conventional and unconventional superconductors like that in ref. 4 for ΔC, only in the current case discussing the condensation energy. This is done in the spirit[7] that it is possible to arrive at phenomenological insights even when a more microscopic understanding is lacking – as was the case when the ΔC ∝ T$_c^{-3}$ correlation was first observed by BNC.[2] Certainly an investigation of the condensation energy of superconductors in general seeking to find the same sort of global correlation as found for ΔC in IBS is of interest.

The first question posed is: do the IBS have an intrinsically different behavior vs BCS superconductors, of U with T$_c$, like found[2-5] for ΔC vs T$_c$? Second, is there, like found for ΔC, a power law U ∝ T$_c^\alpha$ although – just as previously[6-8] for ΔC - there is at present no theoretical basis to expect one? Third, is the condensation energy in s-wave, electron-phonon coupled BCS superconductors comparable with that in unconventional superconductors like the cuprates or the heavy Fermions? Specifically, how does U for the optimally doped cuprates, the heavy Fermion superconductors, MgB$_2$ (a two-band BCS superconductor), Sr$_2$RuO$_4$, Li$_{0.1}$ZrNCl, and the organic superconductor κ-(BEDT-TTF)$_2$Cu(NCS)$_2$ compare to U(T$_c$) for BCS and IBS? If they are different, is there a scaling which creates a common behavior, and can this scaling motivate future theoretical work for better understanding superconductivity?

## II. Experimental

As discussed in refs. 3 and 21 (see also ref. 22), samples of Ba(Fe$_{1-x}$Co$_x$)$_2$As$_2$ were prepared by annealing self-flux-grown single crystals, with the nominal concentration x ranging from 0.055 to 0.15 and 11.6 K < T$_c$ < 24.6 K. The optimal single-temperature annealing procedure was determined[3,21] to be 2 weeks at 700 °C, in the presence of an As vapor source, with extended slow annealing down to 600 °C carried out in one sample (x=0.0766, ref. 22) for strain relief and further sample optimization. The low temperature specific heat was measured by established techniques.[4,23] Additionally for the iron based superconductors, the specific heats for FeSe[24] – T$_c$=8.1 K, FeTe$_{0.57}$Se$_{0.43}$[25] – T$_c$=14.2 K, LiFeAs[26] – T$_c$=14.8 K, Ba$_{0.65}$Na$_{0.35}$Fe$_2$As$_2$[27] – T$_c$=29.4 K, and Ba$_{0.6}$K$_{0.4}$Fe$_2$As$_2$[28] – T$_c$=36.5 K were found in the literature. The specific heats of all the BCS elements with T$_c$>1.4 K (Ta[10], Re[29] – T$_c$=1.7 K, Tl[30] – T$_c$=2.38 K, In[31] – T$_c$=3.405 K, Sn[32] - T$_c$=3.70 K, α-La[33] - T$_c$=4.9 K, V[34] – T$_c$=5.38 K, β-La[33] – T$_c$=6.0 K, Tc[35] – T$_c$=7.86 K, Nb[36] – T$_c$=9.2 K) were taken from the literature, where – due to the rather low upper critical fields – normal state data down to low temperatures are also readily available, with the exception of radioactive technetium, Tc, where an extrapolation[35] was used. The thermodynamic critical field, H$_{c0}$, values[37] for Hg - T$_c$=4.15 K, 411 G, and for Pb – T$_c$=7.2 K, 803 G, and for all the superconducting elements with T$_c$<1.4 K where sufficient specific heat data were lacking, were used to calculate U from $H_{c0}^2/8\pi$ as discussed above. The specific heats of the BCS A15 structure superconductors (non-transforming $V_3Si$[38] – $T_c$=16.6 K, $Nb_3Sn$[39] – $T_c$=17.8 K, and $Nb_3Ge$[40] – $T_c$=21.8 K) were also found in the literature, with normal state data down to 5 K in a 19 T applied field available for $V_3Si$ and down to 10 K in a 16 T applied field for $Nb_3Sn$. An extrapolation[40] of normal state data from above $T_c$ was used for $Nb_3Ge$. The optimally doped cuprate superconductors chosen (and for which condensation energies were available) were: $Bi_{1.74}Sr_{1.88}Pb_{0.38}CuO_6$, $T_c$=9.4 K[41], BiSCCO 2212 $T_c$=83 K[42], $Y_{0.8}Ca_{0.2}Ba_2Cu_3O_7$ $T_c$=83 K[42], and YBCO $T_c$=93 K[18]. The heavy Fermion superconductors chosen were $CeIrIn_5$, $T_c$=0.4 K[43], $UPt_3$, $T_c$=0.54 K[44]., $CeCu_2Si_2$, $T_c$=0.65 K[45], $CePt_3Si$, $T_c$=0.7 K[46], $UBe_{13}$, $T_c$=0.92 K[47], $UNi_2Al_3$, $T_c$=1.0 K[48], $URu_2Si_2$, $T_c$=1.3 K[49], $UPd_2Al_3$, $T_c$=2.0 K[50], $CeCoIn_5$, $T_c$=2.3 K[47], $NpPd_5Al_2$, $T_c$=4.9 K[51], $PuRhGa_5$, $T_c$=9 K[52], and $PuCoGa_5$, $T_c$=18.5 K[53]. Furthermore, the two-band BCS superconductor $MgB_2$[54] and the unconventional superconductors $Sr_2RuO_4$[55], $Li_{0.1}ZrNCl$[56], and $\kappa$-$(BEDT\text{-}TTF)_2Cu(NCS)_2$[57] are also included. Although by no means exhaustive, this choice of 60 different superconductors should be large enough to indicate global trends.

In general – unlike the case for the superconducting elements - normal state data down to low temperatures for the $Ba(Fe_{1-x}Co_x)_2As_2$ samples discussed here do not exist, particularly for the higher $T_c$ compositions $0.07 \leq x \leq 0.0105$. This is due to the rather high[58] upper critical fields, over 35 T. Thus, in order to calculate both the normal state and the superconducting state electronic entropy as a function of temperature, $S_n^e(T)$ and $S_s^e(T)$, for the $Ba(Fe_{1-x}Co_x)_2As_2$ annealed single crystal samples and thus the condensation energy as sketched above in Fig. 1 for Ta, an accurate estimate of the phonon contribution to the specific heat must be made. Unlike in A15 $Nb_3Ge$, where an extrapolation[40] of the normal state data above $T_c$ is simplified by the apparent lack of anharmonic terms in the lattice specific heat in that temperature range, the Co-doped $BaFe_2As_2$ samples are known[59-60] to have both a $T^3$ and an anharmonic $T^5$ term in the lattice specific heat. However, as done in ref. 59, we can use the lattice specific heat of an overdoped, non-superconducting sample of $Ba(Fe_{1-x}Co_x)_2As_2$ to approximate that of the superconducting, lower Co-doped samples. This is because the lattice specific heat (e. g. in the Debye model) is a function of the molar mass of the constituent ions and Co and Fe are very similar in mass resulting in a difference of molar mass between optimally doped and overdoped $Ba(Fe_{1-x}Co_x)_2As_2$ of only 0.12 %. This is the approach used by Hardy et al.[59]. A similar approach was used by Gofryk et al.[60] in their studies of the specific heat of superconducting $Ba(Fe_{1-x}Co_x)_2As_2$ where they used the lattice specific heat of the parent compound, $BaFe_2As_2$. Both methods of approximating the lattice specific heat of the $Ba(Fe_{1-x}Co_x)_2As_2$ compounds for compositions between undoped and overdoped give very similar results.

The calculated values for U as discussed above with Fig. 1 and in the text are all tabulated in Table 1.

### III. Results and Discussion

Before we discuss the full panoply of condensation energy values for all the superconductors, we first concentrate on comparing just the IBS and BCS materials,

$T_c > 1.4$ K, due to the already discovered difference between their respective $\Delta C$ vs $T_c$ behaviors. Shown in Fig. 2 is the superconducting condensation energy, U, vs $T_c$ on a log-log plot for all the superconducting elements with $T_c > 1.4$ K (11 elements in all), 3 A15 superconductors (in red) vs 6 compositions of annealed single crystals of $Ba(Fe_{1-x}Co_x)_2As_2$ plus five other IBS (in black). The specific heat $\gamma$ of the $Ba(Fe_{1-x}Co_x)_2As_2$ samples was fixed by requiring that $S_n^e(T_c)$ (which as discussed with Fig. 1 is just $\gamma*T_c$) be equal to the calculated $S_s^e(T_c)$ (= the integral from 0 to $T_c$ of the difference between the *measured* superconducting specific heat minus the fitted lattice specific heat divided by temperature.) If the lattice specific heat of undoped $BaFe_2As_2$ from ref. 60 is instead used to calculate the electronic entropies for the $Ba(Fe_{1-x}Co_x)_2As_2$ samples, the slope of $\ln(U)$ vs $\ln(T_c)$ for all the IBS shown in Fig. 2 decreases by ~ 0.1. (The use in ref. 60 of the specific heat of the undoped $BaFe_2As_2$ for lattice subtraction may give slightly different results due to the presence of a $T^3$ magnon contribution to the specific heat from the 140 K spin density wave transition.)

It should be stressed for the BCS elements and compounds that the examples chosen cover a large range of coupling strength, from weak ($\lambda=0.46$ for[61] Re, $T_c=1.71$ K) to strong coupled ($\lambda=1.12$ for[61] Pb, $T_c=7.19$ K; $\lambda=1.6$ for[62] $Nb_3Sn$, $T_c=17.8$ K.) Similarly, the examples chosen for the IBS cover a broad range of both $T_c$ and structure.

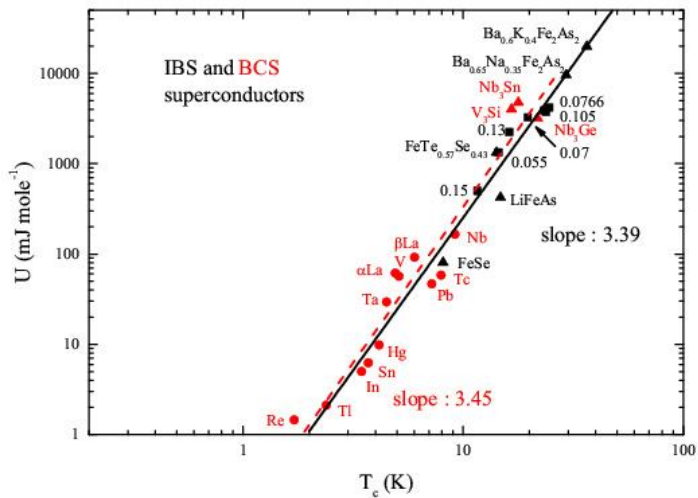

Fig. 2 (color online): The condensation energy for 11 superconducting elements, with two allotropes for La, (red solid circles) and A15 $V_3Si$, $Nb_3Sn$, and $Nb_3Ge$ (red solid triangles), all BCS electron-phonon coupled superconductors calculated as discussed in the text and with Fig. 1. In black (solid squares) is shown the superconducting condensation energy for 6 different superconducting compositions of annealed single crystal $Ba(Fe_{1-x}Co_x)_2As_2$, where the lattice specific heat of an overdoped, non-superconducting sample of $Ba(Fe_{1-x}Co_x)_2As_2$ from ref. 59 was used, plus five other IBS (black solid triangles).

The result of these calculations of U for the electron-phonon coupled superconductors and the IBS are seen in Fig. 2 to give a clear (negative) answer to the question: does U as a function of $T_c$ for these two classes of superconductors – at least in the range of $T_c > 1.4$ K - show an intrinsic difference, as found[4] for $\Delta C$ vs $T_c$? Although the slopes, $\ln(U)$ vs $\ln(T_c)$, of the two sets of superconductors are not exactly equal, the two slope values (3.45 for BCS and 3.39 for IBS) are quite comparable[63,64]. More importantly, the *magnitudes* of the values of U calculated for the two classes of superconductor shown in Fig. 2 are essentially the same vs $T_c$, in <u>strong contrast</u> to the plots[4] of $\ln(\Delta C)$ vs $\ln(T_c)$ where in general $\Delta C$ for the BCS superconductors is much larger (e. g. almost two orders of magnitude at $T_c=4$ K) than that for the IBS. It should be stressed that we are not claiming that U for the IBS and BCS superconductors necessarily can be *precisely* calculated from a simple power law of $T_c$ in this $T_c$ range, that $U_{IBS} = 0.10T_c^{3.39}$, nor that there must be some intrinsic explicit theory underlying the fitted power law. Clearly, the result of Fig. 2 is a phenomenological one pointing out a general trend of U with $T_c$ in the $T_c$ range 1.7 – 36.5 K, with the kernel of truth that the magnitudes of U in both systems, IBS and BCS, are similar – in direct contradiction to the results[4] for $\Delta C$.

This result of $U_{BCS} \approx U_{IBS}$ does hold implications for the theoretical efforts[6-8] for trying to understand the difference between $\Delta C$ vs $T_c$ in the two classes of materials. Specifically, the BCS superconductors gain their condensation energy out of a Fermi liquid normal state. Therefore, if the theory of Zaanen[7] were correct that $\Delta C \propto T_c^{-3}$ in IBS comes from condensation out of a quantum critical normal state, then one would expect a much different U for the Fermi liquid BCS superconductors. Similary, Kogan's model[6] of strong pair breaking for the IBS should also give a much different result that shown in Fig. 2 for well annealed elemental BCS and A15 compound superconductors. The applicability of the multi-band theory of ref. 8 to the present results is currently being calculated, with initial results[65] indicating $U \sim T_c^3$.

Let us now consider the rest of the BCS elements (with $T_c<1.4$ K) with U calculated from their thermodynamic critical fields, $H_{c0}$, which for these type I superconductors is just equal to the upper critical field, $H_{c1}$. (We stress, as already discussed, that the two methods (via specific heat or via $H_{c0}$) for determining U are equivalent and give comparable results.) As shown in Fig. 3, the weaker coupled elements ($\lambda$ for the elements in Fig. 3 is[61], with the

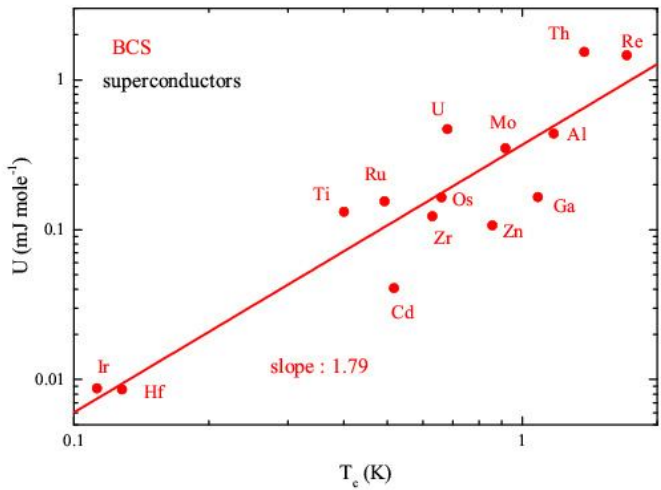

Fig. 3 (color online) Condensation energy U vs $T_c$ on a log-log plot for 13 superconducting elements, $T_c<1.4$ K, plus Re ($T_c=1.71$ K) to provide continuity with Fig. 2.

exclusion of Re, $\lambda=0.46$, between 0.34 to 0.41, while for[61] the elements besides Re in Fig. 2, $\lambda \geq 0.60$) show a.) a large amount of scatter in their U values as a function of $T_c$ and b.) exhibit a *much slower* rise of U with increasing $T_c$ than found for the $T_c>1.4$ K elements in

Fig. 2. (If just the $T_c>1.4$ K elements in Fig. 2, without the three A15 compounds, are fit to $U \sim T_c^\alpha$, the result for $\alpha$ is 2.98). Phenomenologically, the comparison between Figs. 2 and 3 indicates that the superconducting condensation energy in electron-phonon coupled superconductors grows much faster with $T_c$ for medium to strong coupling (Fig. 2), compared to the weak coupling materials in Fig. 3 where BCS theory predicts $U \propto N(0)\Delta^2$. Since in BCS theory $2\Delta/k_B T_c = 3.52$, $U_{BCS}$ should vary as $N(0)T_c^2$.

Now, let us address the question of how these comparisons between U values for IBS and BCS superconductors extend to other superconductors, see Fig. 4. Clearly, neither

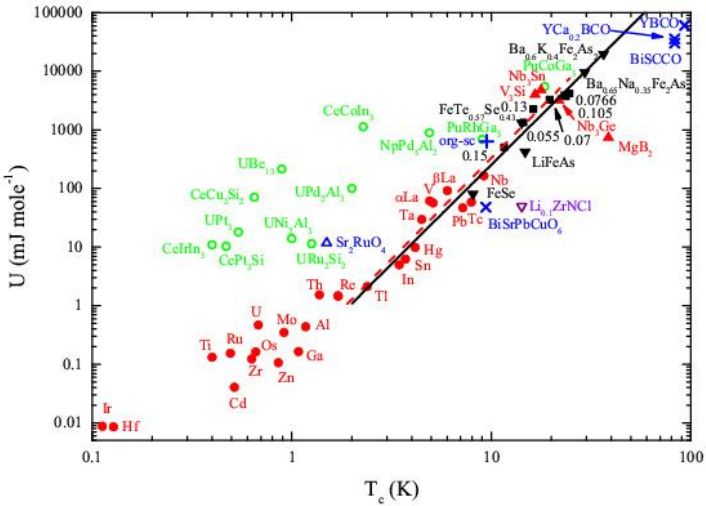

Fig. 4 (color online) Superconducting condensation energy U vs $T_c$ plotted on a log-log scale for heavy Fermion (green), BCS elements (red circles) and A15 compounds as well as $MgB_2$ (red triangles), optimally doped cuprates (blue x's), plus the unconventional superconductors $Sr_2RuO_4$ (blue triangle), $Li_{0.1}ZrNCl$ (inverted triangle) and the organic superconductor (represented by a +) $\kappa$-(BEDT-TTF)$_2$Cu(NCS)$_2$.

the heavy Fermions, nor the optimally doped cuprates, nor the two-band BCS superconductor $MgB_2$, nor the unconventional superconductors $Sr_2RuO_4$, $Li_{0.1}ZrNCl$ and $\kappa$-(BEDT-TTF)$_2$Cu(NCS)$_2$, come anywhere close to agreeing in magnitude with the IBS/BCS behavior found above in Fig. 2. Although the four cuprate points could approximately be taken to lie on a parallel line significantly below (~factor 8) the IBS/BCS trend, the heavy Fermion superconductor U vs $T_c$ exhibit a large amount of scatter and lie very much higher (~2 orders of magnitude) in U at a given $T_c$.

As discussed above in the Introduction, this is not unexpected since when the superconducting energy gap opens in the normal state quasiparticle spectrum, the amount of energy gained by Cooper pair condensation into the superconducting state depends on both the coupling strength, $\lambda$, and the bare density of states $N(0)$ (just as qualitatively given in the weak coupling BCS relation $U \propto N(0)\Delta^2$). Heavy Fermions are well known for being strongly coupled, and the variation in the normal state $\gamma$, $\propto N(0)(1+\lambda)$, is quite large. In the examples shown in Fig. 4, (see Table 1), $\gamma$ varies more than a factor of 20 in the heavy Fermion systems plotted.

Although the 25% decrease in U of, e. g., Pb from the average BCS trend has long since been shown[19] by calculation to be due to strong coupling, and although the decrease of U in the cuprates has also been calculated[15] as due to strong coupling effects, it is undeniable that heavy Fermion superconductors, as a class, exhibit convincing evidence[66-67] of strong coupling as well. Thus, when theory begins to address the condensation energy in heavy Fermions, it must – in agreement with Fig. 4 – find that strong coupling is, in the case of the heavy Fermions, consistent with an *enhanced* U.

     Absent any theory extant for trying to bring order to the extremely wide range of U vs $T_c$ exhibited in Fig. 4, we are left with the desire to find a phenomenological correlation (just as was done[2] in the IBS for $\Delta C \sim T_c^3$) to inspire and focus further experimental and theoretical work. Is there a scaling procedure that condenses the disparate U vs $T_c$ data in Fig. 4 onto one line vs $T_c$? There are two metrics for judging the size of the condensation energy in weak coupling BCS theory: $U \propto N(0)\Delta^2$ or[16] $U \propto \Delta C^* T_c/6.08$. (Obviously, weak coupling BCS theory does not apply to most of the superconductors in Fig. 4, but these two simple relationships offer possible scaling to initially try.) The size of the normal state electronic density of states, N(0), can be estimated by the normal state specific heat $\gamma$, $\propto N(0)(1+\lambda)$. Thus, for example $MgB_2$ has – for the size of its $T_c$ - the lowest lying U in Fig. 4 and a very low density of states estimated from either its low $\gamma$ or $\Delta C$ (see Table 1). The heavy Fermion superconductors, on the other hand, have extremely large U values vs $T_c$ and accompanying large $\Delta C$ and $\gamma$ values (Table 1). For example consider that $\Delta C/T_c$ for $CeCoIn_5$ is over 1500 mJ/molK$^2$ – a truly enormous value.

     Therefore, in Fig. 5 we present $U/\gamma$ vs $T_c$ on a log-log plot. For completeness, we also present $U/[\Delta C/T_c]$ vs $T_c$ in Fig. 6. (These data include only four[68-70] of the 13 low $T_c$ BCS elements for which literature values of $\Delta C/T_c$, listed in Table 1, were found.) Since the ratio $[\Delta C/T_c]/\gamma$ (or $\Delta C/\gamma T_c$) is experimentally roughly 1.5±0.5 (1.43 in BCS weak coupling theory) for most superconductors, the approximately identical results in Figs. 5 and 6 should not be a surprise. We now discuss as representative (and more complete) the $U/\gamma$ vs $T_c$ results in Fig. 5.)

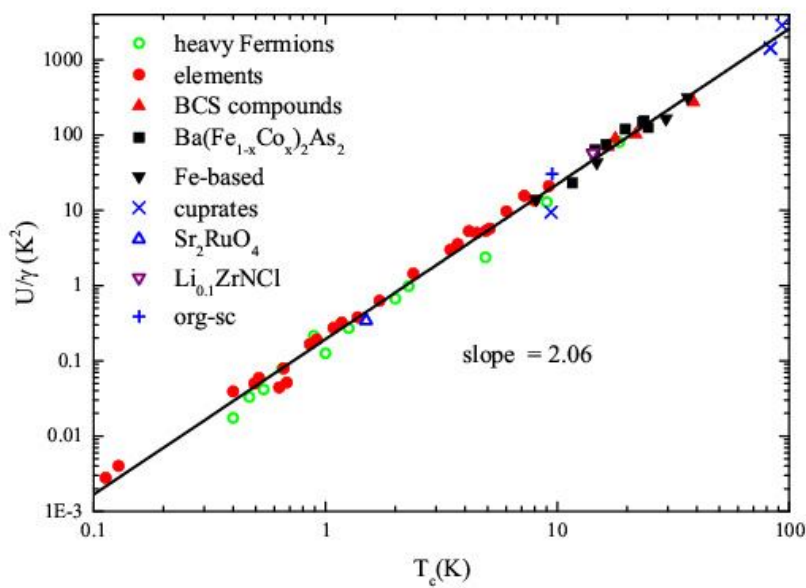

Fig. 5 (color online): Condensation energy U divided by the specific heat $\gamma$ (see Table 1 for values) vs $T_c$ for a wide range of superconductors. There is relatively little scatter about the best-fit line, which covers 6 orders of magnitude for $U/\gamma$ and almost 3 orders of magnitude for $T_c$.

The wide disparity of U with $T_c$ data presented in Fig. 4, with s-, p- ($Sr_2RuO_4$), d- (cuprates), and f-wave ($UPt_3$) pairing symmetry superconductors, collapses rather well onto one line in both figures 5 and 6. Thus, the ratio of the energy gained in a metal by condensing into the superconducting state to the size of the normal state electronic specific heat at $T_c$ ($\propto \gamma$) or to $\Delta C$ is seen to follow – with some scatter – a universal behavior *over 6 orders of magnitude in U* with $T_c$: $U/\gamma$ and $U/[\Delta C/T_c] \approx 0.2 T_c^{\approx 2}$. Therefore, the large scatter in U vs $T_c$ for the heavy Fermion superconductors noted in Fig. 4 is now seen to be just a function of the large spread (from 50 to over 1100 $mJ/molK^2$) in $\gamma$ values.

As a further example of the utility of Fig. 5, consider that the IBS superconductor $Ba_{0.6}K_{0.4}Fe_2As_2$, $T_c \sim 37$ K — with pairing theorized to be due to exchange of spin fluctuations – has (see Table 1) a condensation energy of ~20,000 mJ/mol, while U for the electron-phonon coupled $MgB_2$ at around the same $T_c$ is only ~ 700 mJ/mol, almost a factor of 30 different. However, the two superconductors' *scaled* condensation energies come within 15% of one another, independent of coupling strength or pairing mechanism.

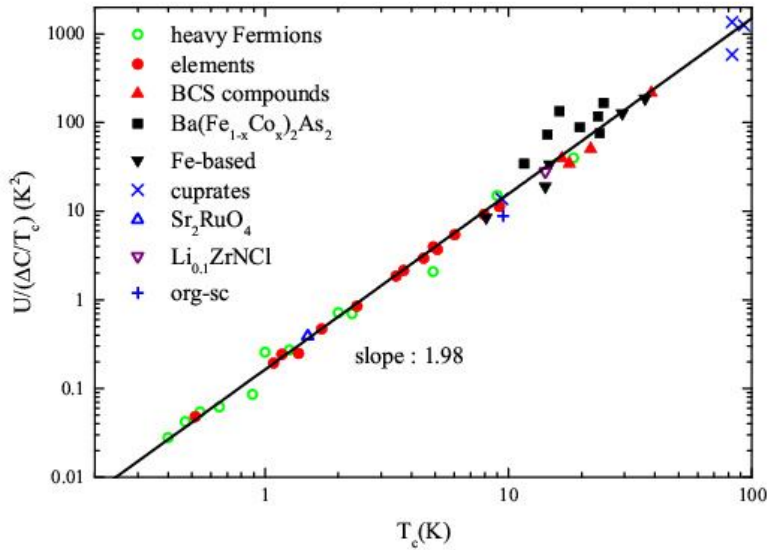

Fig. 6 (color online) $U/[\Delta C/T_c]$ vs $T_c$ on a log-log scale for the same superconductors (with the exception of nine of the low $T_c$ BCS elements for which no $\Delta C/T_c$ data were found in the literature) presented in Figs. 4 and 5 and discussed in the text. The scatter above the general trend in the Co-doped 122 IBS superconductors (black squares) is thought to be due to rather broad transitions[21-22,59] which cause $\Delta C/T_c$ to be underestimated.

Clearly, although weak coupling BCS theory supplied two possibilities for scaling, neither relationship for U ($\propto N(0)\Delta^2$ or $\Delta C^*T_c$) comes close to matching the global trend over all coupling strengths displayed in Figs. 5 and 6. Taking $\Delta \propto T_c$ would give in the first instance $U/N(0) \propto T_c^2$, which is at least reminiscent of Fig. 5, except for the fact that $T_c$ and $N(0)$ (and therefore $\gamma$) are related (see e. g. ref. 61). The second relationship[16], $U \propto \Delta C^*T_c$ - at least for the BCS or IBS materials where a phenomenological trend for $\Delta C$ as a function of $T_c$ ($T_c^{1.8}$ or $T_c^{2.5-3}$ respectively) is known – also fails for the result shown in Fig. 6.

Just as with the global correlation[2-4] in the IBS that $\Delta C/T_c \sim T_c^{1.5-2}$ is only approximate, with differences due[21], for example, to sample quality (annealing), it is important to note that the results given here for the condensation energy also have their limits.  For example, in regards to the cuprates, due to the strong influence of the pseudogap, we have considered only optimally doped samples.  However, in the cuprates, Loram et al.[71] studied U and $\Delta C/T_c$ in $Y_{0.8}Ca_{0.2}Ba_2Cu_3O_{7-\delta}$ with $T_c$ varying between 50 and 83 K, which included a large region of underdoping with pseudogap behavior.   They noted that $U/[\Delta C/T_c]$ was approximately a <u>constant</u> (although with variations in the ratio as large as 40%), independent of $T_c$ and in contradiction to the overall trend found in the present work.

Interestingly, in another work[42] on $Y_{0.8}Ca_{0.2}Ba_2Cu_3O_{7-\delta}$ in the overdoped (no pseudogap) regime, Loram et al. found that $U/\gamma T_c^2$ was approximately constant for six samples with $\delta < 0.26$, which is exactly the result of Fig. 5 of the present work for all the various superconductors studied here.

## IV. Conclusions

For the IBS and BCS superconductors, over a limited (but greater than a decade) $T_c$ range (1.7 – 36.5 K), we have found a universal behavior of U with just the superconducting transition temperature $T_c$, $U \approx 0.1 T_c^{3.4\pm0.2}$.  This result for medium to strong coupled IBS and BCS superconductors may have worthwhile input to further theoretical understanding, since it seems to clearly distinguish itself fairly rapidly as a function of $T_c$ and $\lambda$ from the weak coupled BCS-element regime, $T_c<1.4$ K.  As made abundantly clear:  a.) in Fig. 4 $U \approx 0.1 T_c^{3.4\pm0.2}$ fails to describe the wider variety of superconductors discussed herein and b.) Fig. 3 shows that the higher $T_c$, more strongly coupled behavior in Fig. 2 also does not match the weak coupled BCS elemental behavior, $T_c<1.4$ K.  However, scaling the condensation energy of a superconductor by $\gamma$ or $\Delta C/T_c$ *does* provide a universal behavior for all superconductors considered here.  It is worth stressing that we attempted to include examples of a broad range of superconducting classes, with the pseudogap underdoped cuprates an exception since, see e. g. ref. 18, in the pseudogap regime $\gamma$ and $T_c$ stay relatively constant while U is strongly suppressed.  Thus, with that exception, we could not find a superconductor that did not follow $U/\gamma \approx 0.2 T_c^2$  The question this result raises is:  is there a mechanism to theoretically justify the result $U/\gamma$ (or $U/[\Delta C/T_c]$) $\propto T_c^2$, which extends *far beyond* the weak coupling regime and applies equally to conventional and unconventional superconductors?

Acknowledgements:  The authors gratefully acknowledge helpful discussions with Y. Bang.  Work at Florida performed under the auspices of Basic Energy Sciences, US Department of Energy, contract no. DE-FG02-86ER45268.

______________________________________________________________________________
TABLE I
Parameters for Various Superconductors: heavy Fermions (green), $Sr_2RuO_4$ (blue), BCS elements and A15 compounds together with $MgB_2$ (red), iron based (black), optimally doped cuprates (blue), $Li_{0.1}ZrNCl$, and the organic superconductor κ-$(BEDT-TTF)_2Cu(NCS)_2$. The values for U and $\Delta C/T_c$ for the iron based superconductors are scaled to be for a 100% superconducting sample (due to the significant residual gamma values at low temperatures) by the factor $\gamma_{normal}/(\gamma_{normal} - \gamma_{residual})$, see ref. 21. The γ values (column 5) given for the heavy Fermion compounds are from extrapolations of normal state data to T=0 that give agreement between $S_n(T_c)$ and $S_{sc}(T_c)$. For the $T_c<1.4$ K BCS elements, the second column lists $H_{c0}$ (units of G) in black from ref. 37 instead of $\Delta C/T_c$ except for $Al^{68}$, $Cd^{69}$, $Ga^{69}$, and $Th^{70}$ for which both parameters are listed separated by '/'. γ values for all the elements are from ref. 23.

| Name | $\Delta C/T_c$ (mJ/molK²) | $T_c$(K) | U(mJ/mol) | γ(mJ/molK²) | U/γ (K²) | U/($\Delta C/Tc$) (K²) | reference |
|---|---|---|---|---|---|---|---|
| CeIrIn₅ | 390 | 0.4 | 10.9 | 625 | 0.01744 | 0.02795 | 43 |
| CePt₃Si | 244 | 0.47 | 10.3 | 312.5 | 0.03296 | 0.04221 | 46 |
| UPt₃ | 330 | 0.54 | 18 | 432 | 0.04167 | 0.05455 | 44 |
| CeCu₂Si₂ | 1150 | 0.65 | 71 | 900 | 0.07889 | 0.06174 | 45 |
| UBe₁₃ | 2500 | 0.89 | 214 | 994 | 0.2153 | 0.0856 | 47 |
| UNi₂Al₃ | 54 | 1.0 | 13.9 | 110 | 0.1264 | 0.2574 | 48 |
| URu₂Si₂ | 41.8 | 1.26 | 11.4 | 42.1 | 0.2708 | 0.2727 | 49 |
| UPd₂Al₃ | 140 | 2.0 | 100 | 150 | 0.6667 | 0.7143 | 50 |
| CeCoIn₅ | 1625 | 2.28 | 1136 | 1150 | 0.9878 | 0.6991 | 47 |
| NpPd₅Al₂ | 430. | 4.9 | 893 | 374 | 2.39 | 2.08 | 51 |
| PuRhGa₅ | 46. | 9.0 | 692 | 53.5 | 12.9 | 15.0 | 52 |
| PuCoGa₅ | 137 | 18.5 | 5475 | 68 | 80.5 | 40.0 | 53 |
| | | | | | | | |
| Sr₂RuO₄ | 30 | 1.5 | 11.7 | 34 | 0.3441 | 0.39 | 55 |
| | | | | | | | |
| MgB₂ | 3.4 | 38.7 | 738 | 2.69 | 274.3 | 217.1 | 54 |
| | | | | | | | |
| Al $H_c(0)$=104.9/1.81 | | 1.175 | 0.438 | 1.36 | 0.322 | 0.242 | 37, 23, 68 |
| Cd $H_c(0)$=28.05/0.845 | | 0.517 | 0.0407 | 0.687 | 0.0592 | 0.0482 | 37, 23, 69 |
| Ga $H_c(0)$=59.3/0.85 | | 1.0833 | 0.1651 | 0.6 | 0.275 | 0.194 | 37, 23, 69 |
| Hf $H_c(0)$=12.7 | | 0.128 | 0.00860 | 2.15 | 0.004 | --- | 37, 23 |
| Ir $H_c(0)$=16 | | 0.1125 | 0.00874 | 3.14 | 0.00278 | --- | 37, 23 |
| Mo $H_c(0)$=96.86 | | 0.916 | 0.3504 | 1.83 | 0.1915 | --- | 37, 23 |
| Os $H_c(0)$=70 | | 0.66 | 0.1641 | 2.05 | 0.0800 | --- | 37, 23 |
| Ru $H_c(0)$=69 | | 0.493 | 0.1548 | 3.1 | 0.0499 | --- | 37, 23 |
| Th $H_c(0)$=160/6.2 | | 1.374 | 1.535 | 4.06 | 0.3781 | 0.248 | 37, 23, 70 |
| Ti $H_c(0)$=56 | | 0.40 | 0.1316 | 3.36 | 0.0392 | --- | 37, 23 |
| U $H_c(0)$=100 | | 0.68 | 0.4689 | 9.14 | 0.0513 | --- | 37, 23 |
| Zn $H_c(0)$=54.1 | | 0.857 | 0.1068 | 0.64 | 0.1669 | --- | 37, 23 |
| Zr $H_c(0)$=47 | | 0.63 | 0.1232 | 2.77 | 0.0445 | --- | 37, 23 |
| | | | | | | | |
| Re | 3.1 | 1.71 | 1.46 | 2.32 | 0.6293 | 0.471 | 29 |

| Name | ΔC/Tc (mJ/molK²) | Tc(K) | U(mJ/mol) | γ(mJ/molK²) | U/γ (K²) | U/(ΔC/Tc) (K²) | reference |
|---|---|---|---|---|---|---|---|
| Tl | 2.5 | 2.39 | 2.12 | 1.47 | 1.442 | 0.848 | 30 |
| In | 2.7 | 3.46 | 5 | 1.66 | 3.012 | 1.852 | 31 |
| Sn | 2.9 | 3.71 | 6.26 | 1.76 | 3.557 | 2.159 | 32 |
| Hg $H_c(0)$=411 | | 4.15 | 9.82 | 1.85 | 5.308 | -- | 37, 23 |
| Ta | 10 | 4.48 | 29.5 | 5.87 | 5.026 | 2.95 | 10 |
| α La | 15.5 | 4.91 | 61.5 | 11.5 | 5.348 | 3.968 | 33 |
| V | 15.3 | 5.11 | 56.9 | 9.9 | 5.747 | 3.719 | 34 |
| β La | 16.9 | 6.02 | 91.9 | 9.45 | 9.725 | 5.438 | 33 |
| Pb $H_c(0)$=803 | | 7.196 | 46.8 | 2.99 | 15.652 | -- | 37, 23 |
| Tc | 6.3 | 7.95 | 58.3 | 4.3 | 13.56 | 9.254 | 35 |
| Nb | 14.5 | 9.17 | 164.6 | 7.8 | 21.1 | 11.35 | 36 |
| $V_3Si$ | 101 | 16.6 | 4014 | 57.1 | 70.3 | 39.74 | 38 |
| $Nb_3Sn$ | 138 | 17.8 | 4757 | 53.4 | 89.08 | 34.47 | 39 |
| $Nb_3Ge$ | 63 | 21.8 | 3175 | 31 | 102.4 | 50.4 | 40 |

TABLE I. *(Continued).*

| Name | ΔC/Tc (mJ/molK²) | Tc(K) | U(mJ/mol) | γ(mJ/molK²) | U/γ (K²) | U/(ΔC/Tc) (K²) | reference |
|---|---|---|---|---|---|---|---|
| FeSe | 9.4 | 8.11 | 80 | 5.73 | 13.96 | 8.511 | 24 |
| $FeTe_{57}Se_{43}$ | 69.5 | 14.2 | 1328 | 23.3 | 57.0 | 19.11 | 25 |
| LiFeAs | 12.5 | 14.8 | 423 | 9.67 | 43.74 | 33.84 | 26 |
| $Ba_{0.65}Na_{0.35}Fe_2As_2$ | 75 | 29.4 | 9570 | 57.7 | 165.9 | 127.6 | 27 |
| $Ba_{0.6}K_{0.4}Fe_2As_2$ | 106 | 36.5 | 19800 | 62.5 | 316.8 | 186.8 | 28 |
| $Ba(Fe_{1-x}Co_x)_2As_2$ | | | | | | | |
| x= | | | | | | | |
| 0.15 | 14.6 | 11.6 | 502 | 21.6 | 23.24 | 34.38 | 3 |
| 0.105 | 49.3 | 23.7 | 3740 | 23.9 | 156.5 | 75.85 | 3 |
| 0.13 | 16.8 | 16.2 | 2255 | 30.1 | 74.92 | 134.2 | 3 |
| 0.055 | 18.3 | 14.5 | 1340 | 21 | 63.81 | 73.22 | 3 |
| 0.07 | 37 | 19.65 | 3280 | 27.2 | 120.6 | 88.65 | 3 |
| 0.0766(b) (900 C) | 33 | 23.3 | 3891 | 26.3 | 147.9 | 117.9 | 22, this work |
| 0.0766(a) (600 C) | 25 | 24.6 | 4175 | 33.2 | 125.8 | 167 | 22, this work |
| BiSrPbCuO6 | 3.5 | 9.4 | 48 | 5.1 | 9.412 | 13.72 | 41 |
| YCa0.2BCO | 61 | 83 | 35840 | 25 | 1434 | 587.5 | 42 |
| BiSCCO | 22 | 83 | 30000 | 21 | 1429 | 1364 | 42 |
| YBCO | 48 | 93 | 60471 | 21 | 2880 | 1260 | 18 |
| $Li_{0.1}ZrNCl$ | 1.77 | 14.2 | 49.4 | 0.85 | 58.12 | 27.91 | 56 |
| Org-sc | 72.2 | 9.5 | 641 | 21 | 30.52 | 8.878 | 57 |